\documentclass[preprint2]{aastex}

\newcommand \kms {km\thinspace s$^{-1}$}
\newcommand \ms  {m\thinspace s$^{-1}$}

\shorttitle{TESS Planet and False Positive Rates}
\shortauthors{Brown \& Latham}

\begin{document}


\title{Expected Planet and False Positive Detection Rates for the Transiting
Exoplanet Survey Satellite}


\author{Timothy M. Brown}
\affil{Las Cumbres Observatory Global Telescope, Goleta, CA 93117}
\email{tbrown@lcogt.net}
\and
\author{David W. Latham}
\affil{Harvard-Smithsonian Center for Astrophysics, 60 Garden Street,
Cambridge, MA 02138}
\email{dlatham@cfa.harvard.edu}


\begin{abstract}

The proposed Transiting Exoplanet Survey Satellite (TESS) will survey
the entire sky to locate the nearest and brightest transiting
extrasolar planets with orbital periods up to about 36 days.  Here we
estimate the number and kind of astrophysical false positives that
TESS will report, along with the number of extrasolar planets.  These
estimates are then used to size the ground-based follow-up observing
efforts needed to confirm and characterize the planets.  We estimate
that the needed observing resources will be about 1400
telescope-nights of imaging with 0.5m to 1m-class telescopes, 300
telescope-nights with 1m to 2m-class telescopes for the classification
of the host stars and for radial velocity measurements with roughly 1
\kms\ precision, and 380 telescope-nights with 2m to 4m-class
telescopes for radial velocity studies with precision of a few \ms.
Follow-up spectroscopy of the smallest planets discovered by TESS at
the best possible velocity precision will be limited by the number of
telescope nights available on 4m to 10-m class telescopes with
instruments such as HARPS and HIRES, but the pay-off of such efforts
will be the determination of masses for Super Earths with sufficient
accuracy to distinguish rocky desert planets from water worlds.

\end{abstract}




\section{History and Motivation}

Photometric surveys to detect transits by extrasolar planets have
recently become a productive means of locating these objects
\citep[e.g., ][]{koc98,alo04b,bak04,bag07,wal03}.  All such surveys
must deal with astrophysical false positives, i.e., periodic
transit-like decreases in stellar brightness that arise from stars
orbited by other stars, as opposed to stars orbited by planets.  In
most surveys these false positives outnumber those from planetary
transits, often by factors of ten or more.

The proposed Transiting Exoplanet Survey Satellite (TESS) seeks to use
an orbiting array of 6 small-aperture, wide-field CCD cameras (190 mm
focal length f/1.5, with 18-degree square field of view) to survey the
entire sky in a 2-year observing campaign, yielding a comprehensive
list of transiting planets orbiting 2.5 million of the nearest stars.
From a near-equatorial orbit, TESS will be able to observe the whole
sky, and to detect objects
with transit depths as small as $3 \times 10^{-4}$ of the brightness
of the parent star, with periods up to 36 days.  These
sensitivities will allow detection of Neptune-sized planets transiting
Sun-like stars, and Super Earths around stars with radii somewhat
smaller than the Sun's. TESS was selected by NASA for Phase A in May 2008.  

In order to size the ground-based follow-up effort needed to
distinguish true planets from false positives, one must estimate the
rates of false positives that TESS will produce.  This paper describes
how we made such estimates, and what we learned.  The point of
departure for our computations was the simulation code described by
\citet{bro03}, which should be consulted for information about the
computational strategy.  Modifications to the code were needed to make
estimates that were appropriate to TESS; the most important of these
was the need to average the rates for line-of-sight triple star
systems (two stars of which compose an eclipsing binary) over all Galactic
latitudes.  Lesser modifications brought the code up to date in terms
of the period and radius distributions of presently-known extrasolar
planets.

\section{Categories of False Positives}

Only a few physical processes cause most of the light variations that
can masquerade as transits by planet-sized bodies.  Each of these has
its own characteristic distributions of transit depth, duration, and
period, and each is (in principle) distinguishable from planetary
transits by one or more kinds of follow-up observations.  Examples of
the methods typically used for ground-based transit surveys may be
found in, e.g., \citet{odo07} and \citet{lat08}.  The ones that are
treated in this paper are described below.


Eclipsing binaries (EBs) in which the brighter object is a
main-sequence star generate eclipses that have typical depths of tens
of percent; these are usually so deep that confusion with transiting
planets does not occur.  But in a small fraction of cases, the
eclipses are grazing, and eclipse depths can be a few percent or less.
These grazing eclipses of stars can most easily be distinguished from
those of planets if the light curve has high enough precision that the
(generally flat-bottomed) eclipse shape generated by a planet can be
distinguished from the (generally V-shaped) eclipses that result from
stellar grazes.  For transit observations with lower signal-to-noise
ratio (S/N), the most powerful method for distinguishing EBs from
planetary transits is to observe the radial velocity variations of the
primary star.  For a stellar-mass companion, the orbital velocities
are typically tens of \kms, while for planetary-mass companions, they
are only hundreds of \ms\ at most.  Thus high-resolution spectroscopy
at modest S/N is an effective way to identify systems where stellar
companions are responsible for the transit-like light curves.  For
example, radial velocities with a precision of 1 \kms\ are good enough
to detect companions down to about 10 Jupiter masses for orbital
periods of a few days around sun-like stars.

Eclipsing binaries in which the primary is a giant star usually have
main-sequence companions, and it is not hard to find such pairs that
yield eclipse depths similar to planetary transits.  However, the
eclipse duration for a dwarf eclipsing a giant tends to be much longer
than for a planet transiting a dwarf, and this provides an effective
way to distinguish between the two configurations.  Furthermore,
giants are typically too large to allow orbits with periods of just a
few days, which is the domain favored by planets transiting sun-like
stars.  Spectroscopy of gravity-sensitive features such as the Mg b
lines in the green is another effective way to identify host stars
that are giants \citep{lat08}.

More difficult diagnostic problems arise if the light from an
eclipsing binary is diluted by that from a brighter third star.  An
important class of these systems consists of an EB that, by chance,
lies nearly along the same line of sight as an apparently brighter
third star (we term these ``line-of-sight triples'').  When the third
star dominates the light that we see, then moderate-sensitivity radial
velocity measurements are likely to show no variability, consistent
with the interpretation that the companion is a planet.  The
likelihood of random superpositions is larger than one might suppose,
because transiting-planet surveys typically use very low spatial
resolution, with stellar image diameters ranging from about 8 arcsec
(for Kepler) to more than 30 arcsec (for most ground-based wide-angle
surveys).  To identify diluted systems of this sort, the easiest
diagnostics involve improving the spatial resolution.  One method that
can be applied to the survey data itself is to use image subtraction
to localize the variability associated with the system's transits.  If
this center-of-variability does not coincide with the position of the
nominal target star, then the variability is likely to arise from a
background eclipsing binary.  Another powerful diagnostic method is to
observe transits with much higher spatial resolution than that which
was used to detect them.  If the system is a line-of-sight triple,
then the relatively large eclipses of the faint EB component will
usually be easily observable if the EB is spatially resolved.  Because
the survey images are so large, seeing-limited resolution will often
be enough to separate these systems.  In extreme cases, however,
adaptive optics (AO) observations or even high-resolution imaging from
above the Earth's atmosphere may be desirable.

Finally, and most difficult, one encounters gravitationally-bound
triple systems (``physical triples'') that behave similarly to the
line-of-sight triples just discussed.  In the bulk of these cases, the
angular separation between the EB system and the third star will be so
small that they will remain spatially unresolved even with AO systems.
In these cases, distinguishing planetary from stellar transits may be
tackled using either photometry or spectroscopy.  A photometric
diagnostic is possible if the primary star of the EB system has a
significantly different color than the third (brightest) star.  If
this is so, then the color of the triple system changes during
eclipses -- something that does not happen for planetary transits
\citep{odo06}.  Lacking a sufficient color difference among the stars
in the system, one must search for the small changes in spectral line
shape that result from the weak line profiles from the EB system
moving across the line profiles from the brightest star.  Analysis of
this sort can be difficult and time-consuming \citep{man05},
especially if the lines of the EB have been broadened due to
synchronization of the stellar rotation with a short orbital period by
tidal mechanisms.

\section{Adaptations for the TESS Calculation}

The programs used here for estimating planet and false positive
detection rates had previously been applied to small-aperture
ground-based surveys and to NASA's Kepler mission \citep{koc04}, in
both cases with reference to restricted fields lying near the Galactic
plane.  Moreover, in the time since the previous calculations,
knowledge of the distribution of exoplanet periods and radii has
improved considerably.  For calculations specific to the circumstances
of TESS, some modifications to the code were necessary.

The simplest adjustment was to set the confusion radius for
line-of-sight triples to a value of 24 arcsec, appropriate to TESS's
16-arcsec pixels and image performance.  However, the probability that
a bright foreground star will have a faint EB in its background
depends not only on the detector's confusion radius, but also on the
areal density of stars.  The latter is of course a strong function of
Galactic latitude, and TESS's all-sky nature thus requires allowance
for this variation.  For this reason we computed probabilities for
line-of-sight triples for each 10-degree increment in the absolute
value of the Galactic latitude, and then averaged the resulting
probability distributions over latitude, weighting by the area of sky
covered by each stripe.

Observations over the past few years have yielded a considerable
amount of information about the distribution of planets with orbital
period, and also (with substantial input from theory, eg.
\citet{ida05}) a schematic picture of the distribution with planetary
radius.  We therefore updated the parameterization used in
\citet{bro03}.  For the current purposes, we took the distributions in
both planetary radius and orbital period to be such that equal
cumulative probabilities are found in equal increments of the log of
radius or period, respectively.  We also took the joint probability
distribution to be simply the product of the individual ones, i.e., we
assumed that planetary radius and orbital period are independent
variables.  These distributions are broadly consistent with what we
know about the real distributions of planetary parameters \citep[e.g.,
][]{mar05,udr07}, but the latter are uncertain enough that we have not
attempted a more sophisticated simulation.  Indeed, one of the goals
of TESS is to improve our knowledge of the rates of occurrence of
planets as a function of planetary mass and length of semimajor axis.

\section{Results}

Given the assumptions outlined above, we computed the expected rates
of planet detections and of the various kinds of false positives.  For
purposes of estimating total rates, we assumed that TESS would image
$2.5 \times 10^6$ stars spanning Sloan $r$ magnitudes between 4.5 and
13.5, and we integrated over orbital periods between 1.0 and 36 days,
transit depths between $10^{-4}$ and 0.04, and transit durations
between 0.02 and 0.5 days.  Counting only cases for which at least 3
transits could be observed,
we arrived at the rates shown in Table 1.
Note that this counting gave equal weight to transits of all depths
down to the minimum of $10^{-4}$.
This signal level is appropriate for the brighter stars that TESS
will observe, but is too optimistic by a factor of 3 or more for stars
near the faint magnitude limit.
As a result, Table 1 overestimates the numbers both of detected planets
and of false positives.
Because of the flat distribution of planets with the log of transit depth,
this over-estimate for planets is probably about 20\%.
By similar reasoning, the over-estimate for line-of-sight triples 
is likely larger than for planets, and the over-estimates for EBs 
and physical triples are likely smaller.

\begin{deluxetable}{ll}
\tablewidth{0pt}
\tablecaption{Transiting Planet and False Positive Yields \label{Table 1}}
\tablehead{
\colhead{Signal Type} & \colhead{Number (3 Transits)}
}
\startdata
Planets & 1687 \\
MS Binaries & 2111 \\
Giant Binaries & 40 \\
Dilute MS Binaries & 7237 \\
Physical Triples & 2080 \\
\enddata
\end{deluxetable}

\subsection{Planets}

The expected yield of planets is about 1700, or roughly one per 1500
stars surveyed.  This number is considerably larger than that for
ground-based surveys, a fact that we attribute mostly to TESS's
superior photometric precision and consequent ability to detect
smaller planets, but also to its regular and uninterrupted observing
window function.  The importance of photometric precision is
illustrated in Figure 1, which shows the marginal probability
distribution of planetary transit depths, and also of various
false-positive sources.  This distribution rises for decreasing
transit depths starting at a relative depth of about 0.01,
corresponding to transits of the largest likely planet (roughly
Jupiter-sized) across the most common (roughly Sun-sized) star seen in
small-aperture surveys.  At shallower transit depths, the probability
density per unit $\log$ depth becomes roughly constant.  This behavior
depends fairly sensitively on the assumed distributions of planetary
radii and orbital semimajor axis, and the correlation between them.
But given our assumptions about these distributions, it is clear that
each increase of some factor in the minimum detectable transit depth
yields about a constant increase in the number of detected planets.

\begin{figure}
\plotone{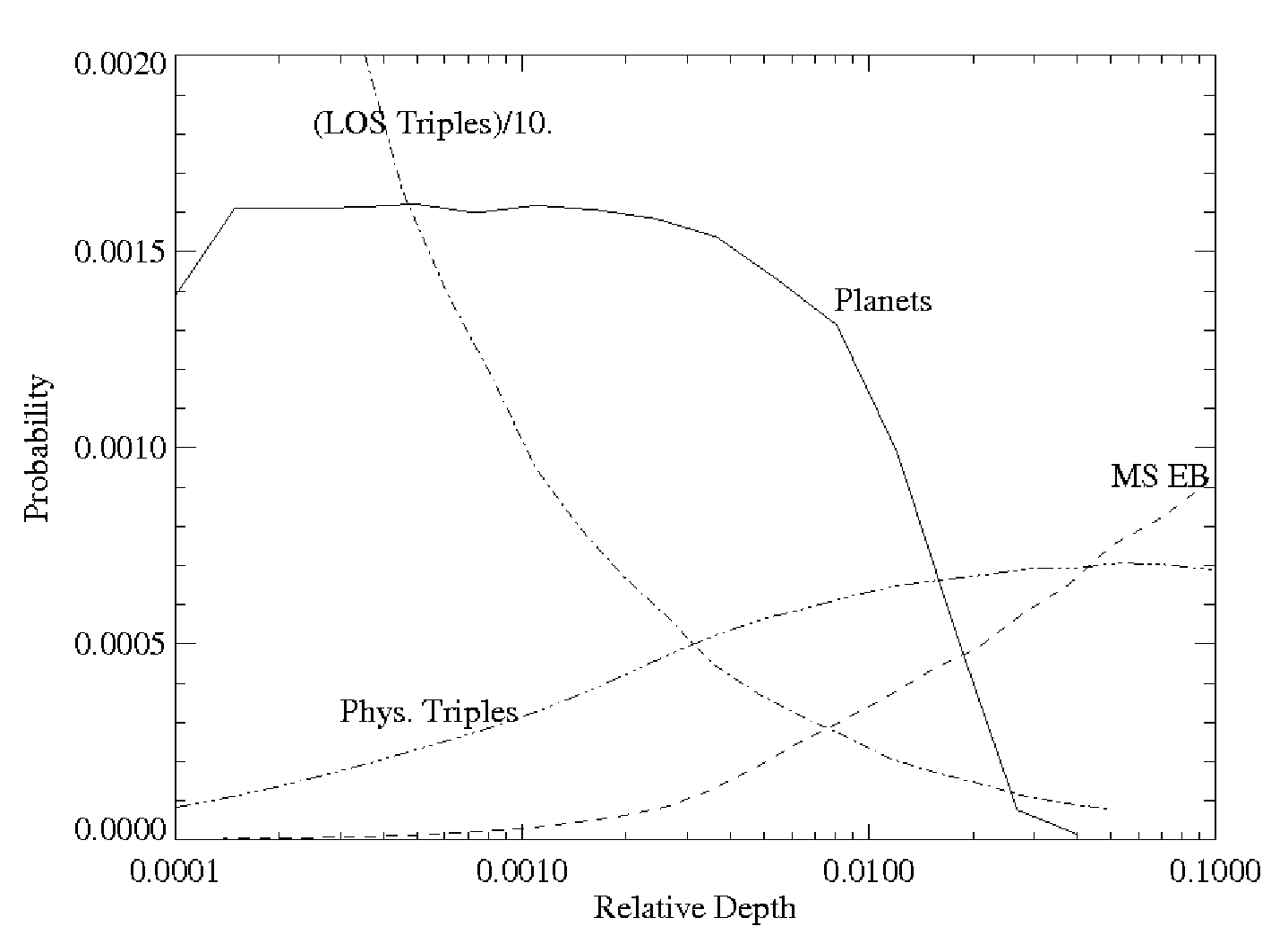}
\caption{Probability density (per unit $\log \Delta$) for the
occurrence of transits with relative depth $\Delta$, for systems in
which the transit is caused by a planet, by a grazing eclipse of a
main-sequence star in an eclipsing binary, and by eclipses in EB
systems that are diluted by light from a third star in a line-of-sight
or a gravitationally-bound (physical) triple star system.  The
probability densities for line-of-sight triples have been divided by
10, to put them on a similar scale to the other curves.  These are the
marginal distributions, i.e., they have been integrated over all the
other variables that describe each type of transit, namely orbital
period, transit duration, and mass of the parent star.}
\end{figure}

The computed distribution of planetary periods rises monotonically
from the longest to the shortest periods considered.  As with
ground-based searches, short-period planets are heavily favored for
discovery, because of the relatively wide range of inclinations over
which they exhibit transits.  Between periods of 1 day and 30 days,
the marginal probability density (per unit $\log$ period) falls by
about a factor of 10.

The distribution of transit durations $d$ is peaked at about 2 hours,
and is nearly Gaussian when plotted against $\log d$, with a full
width at half maximum of about 0.3 in $\log d$ (Figure 2).  
The typical transit
is fairly short compared to the norm for known transiting planets
because small, short-period planets are assumed to be both numerous
and detectable.

\begin{figure}
\plotone{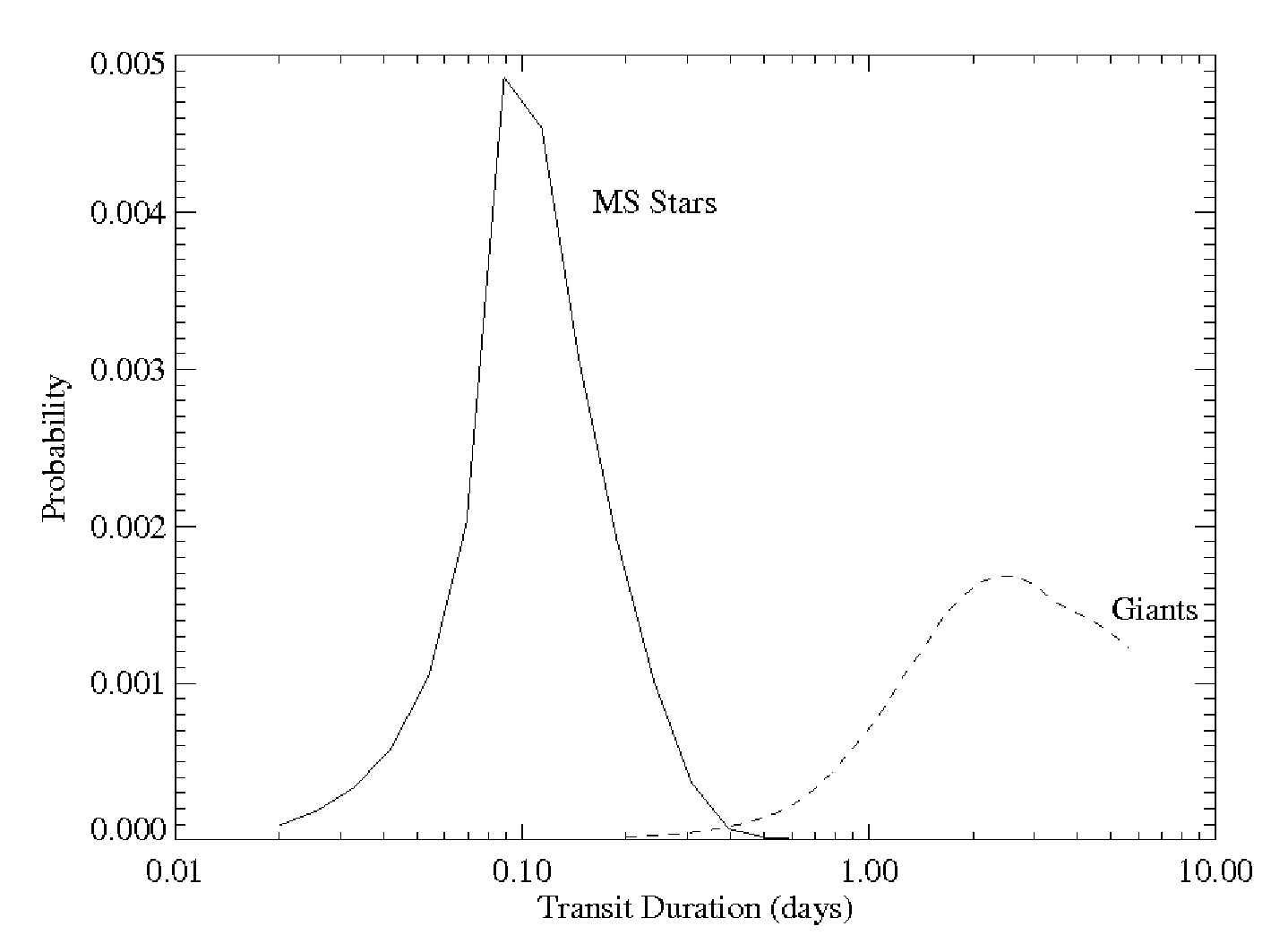}
\caption{Probability density (per unit $\log d$) for the occurrence of
a planet transiting a main-sequence star, or an eclipsing binary with
a giant primary with a transit duration of $d$ days.  As before, these
are marginal distributions, which in this case means that the joint
probability distribution has been integrated over all values of
eclipse depth, orbital period, and primary star mass (for planets) or
($B-V$) color (for giants).}
\end{figure}

\subsection{Main Sequence Eclipsing Binaries}

Ordinary eclipsing binaries in which both components are main-sequence
stars will produce moderately numerous false positives, roughly equal
in number to detected planets.  The expected distribution of transit
depth is, however, quite different from that of planets.  The EB
transit depth distribution peaks at about 25\%, with a long tail to
smaller depths caused by small stars transiting large ones, and by
grazing transits (Figure 1).  Moreover, the geometry of grazing
transits implies that extreme grazes have short durations, and also
that their light curves show distinctive V-shaped eclipses.  Thus,
with high-quality photometry, a large majority of these EB systems can
be rejected as planets from an examination of their light curves.  In
addition, the EBs with short enough periods will show ellipsoidal
variations in their light curves outside eclipses. The remaining cases
will require follow-up medium-precision radial velocity measurements
to distinguish them from planets.

\subsection{Giant Eclipsing Binaries}

Eclipsing binaries in which the brighter component is a giant will be
a relatively unimportant source of false positives.  This is mainly
because the distribution of transit durations peaks at longer
durations (about 2.5d) than is possible for planets in short-period
orbits around roughly Sun-sized stars (Figure 2).  Only a very small
minority of grazing transits will produce light curve dips that are
short enough to cause confusion with planets, and in these, a single
spectrum of the object can reveal that the host star is a giant.
These objects are therefore a negligible driver of the follow-up
effort.

\subsection{Line-of-Sight Triples}

Line-of-sight triples will be the most numerous source of false
positives encountered by TESS, being about 4 times more common than
planetary transits.  TESS's large projected pixel size, and
corresponding large radius of confusion, is the principal reason why
this kind of false positive will be so important.  In the simplest
analysis (which is what we have done here), the distributions of
orbital period and duration should be identical to those for
main-sequence EBs, since line-of-sight triples consist of ordinary EB
systems that happen to lie in nearly the same direction as a third,
brighter star.  According to this picture, the marginal probability
density for transit depth rises monotonically as the depth decreases
(Figure 1), because, in a homogeneous Galaxy, there are always many
more faint EBs per square degree of sky than there are bright ones.
This conclusion cannot be completely true, however, since the Galaxy
is not homogeneous: at high Galactic latitudes, a very faint EB is
almost certain to consist of a pair of red dwarfs, since more luminous
stars would have to be far above the Galactic disk.  Thus, in
principle there should be differences from the naively-calculated
probability densities for small apparent transit depth, due to the
changing nature of the population of eclipsing stars.  We acknowledge
that such subtleties exist, but we have made no effort to model
them here.

A powerful test for line-of-sight triples is to compare the position
on the sky of the putative planet-bearing star with that of the
transit signal.  This can be done first using data from the
space-based transit search itself.  The images from the TESS survey
will have poor spatial resolution but high S/N; with difference-image
analysis it will ordinarily be possible to localize the transit signal
within a small fraction of a pixel.

The remainder will require detailed study of the shape and color of
their light curves, as well as spectroscopy and high-resolution
imaging, to identify the true source of the variations.  Fortunately
at this stage of the follow up, true planets should outnumber false
positives by a considerable margin.

\subsection{Physical Triples}

Physical triple systems will be the most difficult of the false positive
varieties to identify.  They will be roughly as numerous as detected
planets, and in at least some cases their observable characteristics
will closely match those of plausible planetary systems.  Figure 1
shows our estimated marginal probability distribution for transit
depth $\Delta$ for these systems.  The probability density (per unit
$\log \Delta$) for this calculation peaks near transit depth $\Delta =
0.05$, but the distribution is very wide, with a tail that extends
well below TESS's sensitivity limit of $10^{-4}$.  One should
recognize that the details of this distribution depend upon the
poorly-known luminosity distributions within triple star systems;
nevertheless both this model and observational experience
\citep[e.g., ][]{man05} suggest that such triple systems can be a serious
observational challenge.

Fortunately, a fairly large fraction of physical triples will have
characteristics that do allow them to be distinguished from stars with
transiting planets.  Most EB configurations yield a V-shaped eclipse
with such a long duration of ingress and egress that a transiting
planet cannot be the cause.  In many of the remaining cases, the
brightest star in the system is a giant or a hot upper-main-sequence
star, so that a single classification-quality spectrum can show that
the eclipses must be highly diluted.  In cases in which the third
(brightest) star and the EB primary have sufficiently different
temperatures, the transit depth becomes a function of wavelength --
multicolor observations can then reveal the third star.  If the flux
ratio between the third star and the EB primary is not too large, a
composite spectrum will be visible, and one set of lines will show
Doppler shifts varying at the orbital period while the other set does
not.  Finally, if the flux ratio is large, and the stellar rotational
speeds are such as to conceal the EB spectrum within that of the third
star, one may nevertheless infer the third star's presence from
changes in spectral line asymmetries that are in phase with the
eclipse period \citep{man05}.  Verifying this last behavior requires
high-S/N spectroscopy at many orbital phases, and is therefore
expensive in observing time.

\section{Summary:  Follow-up Requirements}

As discussed above, several kinds of ground-based follow-up
observations are feasible and useful, each with its own observing
requirements and each able to identify different kinds of false
positives.  In this section we summarize the role of these
false-positive tests in the context of TESS.
To estimate the amount of follow-up observing time that will be
needed, we must assign values for the fraction of impostors of
different sorts that can be identified using each observing method.
These estimates have various justifications, but sometimes the best
we can offer are guesses, based in experience, but requiring
considerable extrapolation. 
The reader should therefore treat the numerical estimates in the
following paragraphs with caution.

The high-quality TESS photometry itself will allow us to identify some
EBs and physical triples from the variability seen in their
out-of-transit light curves. A crude guess based on our experience with
ground-based surveys is that perhaps 20\% of EBs and 10\% of triples
can be identified in this way.  Comparing the centroid of the transit
signal with that of the apparent host star, as described above, should
be much more effective at sorting out line-of-sight triples, with a
success rate that might reach 90\%.  This is because it should be
possible to determine the relative position of the image centroid in
and out of transit to a small fraction of the image size, roughly the
FWHM divided by the S/N of the transit detection, assumed to be at
least 7. Locating the positional offset during transit to a third of
the image confusion radius should reduce the area allowed for an
accidental alignment by an order of magnitude

Given a good ephemeris for a transiting candidate from the TESS
photometry, searching for line-of-sight triples using seeing-limited
ground-based imaging will also be very effective, and should require
relatively little observing time per target.  The eclipsing object in
these apparent triples will usually be quite faint relative to the
supposed planet host, but for this reason its eclipses must be
relatively deep.  A few observations inside and outside of the
expected transit times thus should suffice to learn whether the
varying star is the supposed planet host star or a faint neighbor.
The large number of candidates for this type of follow-up observation
should allow a multiplex strategy, where several objects can be
followed up simultaneously. Assuming the images are seeing limited, we
estimate that a further 90\% of the line-of-sight triples remaining
from the previous steps can be rejected by these observations, using
about 0.5 hour total time on a 0.5-m class telescope for each target.

A single spectrum of gravity-sensitive features such as the Mg b lines
in the green can immediately identify host stars that are
giants. Luminous giants can be distinguished from dwarfs with
surprisingly poor S/N, as low as 10 to 20 per spectral resolution
element of 10 \kms\ \citep{lat08}. Subgiants, for example near
spectral type G, are more difficult to distinguish from dwarfs,
because their separation in radius/gravity/luminosity is smaller. The
identification of short-period double-lined eclipsing binaries is
often revealed by one or two high-resolution spectra with modest S/N,
thus demonstrating that the transit-like light curves must be due to
grazing eclipses in a system where the secondary is not too much
fainter than the primary. Stars that are rotating much too rapidly to
allow very precise radial velocity measurements can also be revealed
by a single high-resolution spectrum at modest S/N.

We estimate the typical integration time for such spectra to be 10 to
20 minutes with 1m to 2m-class telescopes and modern spectrometers, and
that this will allow discrimination of the roughly 15\% of
main-sequence EBs that show composite spectra, and most of the EBs and
triple systems (both line-of-sight and physical)
where the brightest star is a giant.
Based on the relative numbers of giants and dwarfs among stars in TESS's
magnitude range, we expect that about one-third of triple systems
will be distinguishable in this way.

A few moderate-precision (i.e., 0.5 to 1 km s$^{-1}$) radial velocity
measurements provide almost total rejection of the remaining undiluted
eclipsing binaries.  We assume that each target will require 2-4
radial velocity measurements, and that each of these spectra will also
require 10 to 20 minutes of time on a 1m to 2m-class telescope.
Experience suggests that these data will allow correct identification
of 95\% of EBs, 75\% of Giant EBs, and two-thirds of triples
(both line-of-sight and physical).

Accurate multicolor ground-based light curves provide a means of
identifying a substantial fraction of the remaining diluted EB
systems, namely the ones with deeper transits and big enough color
differences between the EB and diluting star (such light curves are
also necessary for characterizing the properties of genuine planets),
These observations are relatively expensive in observing time because
continuous coverage of full events is needed to achieve the necessary
photometric precision, and one requires time resolution good enough to map out
the shape of ingress and egress.  Transit durations are typically 3
hours; to obtain satisfactory out-of-transit baselines, a typical
transit event requires about 5 hours of telescope time, For relatively
deep transit events (corresponding to giant planets), the precision
achievable for single transit events with 0.5m to 1m-class telescopes
will be adequate. Because of the demand for substantial telescope
time, precise ground-based photometry is usually reserved for targets
that have already been vetted by less demanding methods.

Ground-based photometry of single transit events has rarely achieved a
photometric precision better than 1 mmag in sample times on the order
of a few minutes.  In a few cases larger ground-based telescopes have
been used to push below 1 mmag \citep[e.g., ][]{joh08}), and even better
limiting precision has been demonstrated using a 1m-class telescope to
observe multiple events \citep{win07}. TESS will detect transits that
are an order of magnitude shallower than can routinely be achieved from
the ground.  These will need to be followed up with space-based
resources such as HST and JWST.

In a final step, remaining impostors may be revealed (and true planets
characterized) by precise, multi-epoch, high-resolution Doppler
spectroscopy.  The high spectral resolution and S/N required for these
measurements demands large telescopes, specialized highly stable
spectrometers, and, for faint targets, long exposure times.  To get 4
suitable spectra at the level of a few \ms\ on these objects will
require on the order 2 hours on 2m to 4m-class telescopes, and in some
cases (such as stars with significant rotation or weak lines due to
low metallicity or high temperature) significantly more.  These
observations are expensive and therefore are typically the last to be
done, after all other means of rejecting false positives have been
exhausted.

Last, it is worth noting that a few false positives are likely to
slip through even the exhaustive vetting process just described,
and be counted as planets.
These are most likely to be special kinds of physical triple systems,
in which the EB components are very faint relative to the brightest
star.
These will yield low-amplitude transits with light curves that are
difficult to characterize from the ground.
With a large brightness ratio between the brightest star and the EB
primary, the orbital displacement of the EB spectrum lines will likewise
be difficult to discern.
At the same time, however, the blended line profiles may yield a Doppler
signal that is plausible for a planet orbiting the brightest star.
In short, given the vast number of double and multiple stars in the sky,
one must expect that some of them will succeed in simulating planet-bearing
single stars.
It is difficult to estimate how often this will occur, but given the
special circumstances required, we guess the number will be perhaps
20, or about 1\% of the total sample of planets.

Table 2 shows the expected observing effort implied for the TESS
experiment by the above hierarchy of follow-up observations.  Rows in
the table correspond to the various observational tests that can be
performed.  We assume that these are done in succession, with
viable planet candidates passing to the next test, and
likely false positives being removed from further consideration.  The
first 5 columns show the number of candidate transiting-planet systems
that remain, going into each stage of the process, separated into true
planets and the 4 different kinds of false positive.  Column 6 shows
the number of systems that must be observed at each stage, and column
7 shows the corresponding number of observing hours (and the size of
telescope facilities required).

\begin{deluxetable}{lccccccr}
\tabletypesize{\scriptsize}
\tablecaption{Follow-Up Observing Resources Needed \label{Table 2}}
\tablehead{
\colhead{Observation} & \colhead{Planets} & \colhead{MS EB} &
\colhead{Giant EB} & \colhead{LOS Triple} & \colhead{Phys. Triple} & 
\colhead{\# of Targets} & \colhead{Observing Time}
}
\startdata
Examine TESS & 1687 & 2111 &   40 & 7237 & 2080 & 13155 &    --   \\
Light Curves \\
\\
Seeing-Limited & 1687 & 1689 & 32 & 724 & 1872 & 6004 & 3002h 0.5m-1m \\
Imaging \\
\\
Single Spectrum & 1687 & 1689 & 32 &  72 & 1872 & 5352 & 803h 1m-2m \\
(Classification) \\
\\
6 Radial Velocity & 1687 & 1435 & 10 & 43 & 1123 & 4298 & 2149h 1m-2m \\
Spectra (km s$^{-1}$ \\
\\
High-Quality & 1687 & 72 & 2 & 16 & 404 & 2181 & 10906h 0.5m-1m \\
Light Curve \\
\\
4 High-Quality & 1687 & 29 & 1 & 6 & 162 & 1885 & 3769h 2m-4m \\
RV Spectra \\
\enddata
\end{deluxetable}

The observations that reveal the largest numbers of false positives
are detailed examination of TESS light curves and moderate precision
radial velocity measurements.  In both of these stages the total
number of false positives is reduced by more than a factor of 2.
Intermediate steps are also important, however.  For instance, if
spectroscopy is attempted before most of the line-of-sight triples
have been weeded out, then one is likely to spend much observing time
acquiring spectra of the wrong stars in these systems.  The bulk of
the observing time goes to the last 2 stages of the process, namely
obtaining high-quality light curves and very precise radial velocity
measurements.  By the time the last stage is reached, our experience with
ground-based surveys is that somewhat less than half of the candidates
prove to be planets.
The fraction of true planets should be larger than this for the TESS candidates,
because in the relevant range of transit depth,
the frequency of both main-sequence EBs and physical triples 
is a strongly decreasing function of depth, 
whereas that of planets is almost constant (see Fig. 1).

It is possible to realize some savings in the needed follow-up time by
using existing multicolor photometry or astrometry
(e.g. 2MASS or Tycho, \citet{skr06,hog00}) to identify and discard giants
from the TESS target list.  Systems in which the light is dominated by
a giant are almost always false positives, and there is no need to
spend extra observing time determining to which category of false
positive they belong.

The total observing time that can be saved by a pre-selection that
eliminates giants is fairly modest, however.  The largest contribution
of such false positives comes from eclipsing binaries diluted by
giants.  Such systems should have a rate of occurrence similar to the
fraction of giants in the field. For the
magnitude limit of TESS, this fraction is about one-third.

In any case, most of the systems containing giants are eliminated
early in the sifting process, at or before the step that calls for one
classification-quality spectrum.  These stages are relatively cheap in
observing time.  It is the true planets that must be followed all the
way to the end of the decision tree, incurring high observing time
costs along the way.  Finally, photometry or astrometry suitable for
identifying giants is
unavailable for most of the fainter stars in the TESS sample, and
these are of course by far more numerous than the brighter ones.  For
example, only about 50,000 dwarfs out of TESS's sample of 2.5 million
stars have Hipparcos parallaxes.  Essentially all of the TESS targets
will have 2MASS magnitudes, but with only 2MASS data, the giant-dwarf
distinction is clear only for M stars, which are a small fraction of
the total.

An optimistic estimate of the improvement yielded by pre-selecting
target stars results from assuming that all giant stars can be removed
from the sample before the search for transiting objects begins.  In
this case, we estimate the total observing effort needed for
false-positive rejection will be about 5\% smaller for imaging, and
about 10\% smaller for low-precision radial velocities, than if no
target selection is done.  There would be virtually no change in the
amount of required time for highly precise radial velocities.
Pre-selection of targets is still worthwhile, because the total
amounts of telescope time involved are rather large, and because the
brighter targets (for which pre-selection is most practical) are the
most interesting ones for later follow-up studies, particularly by
JWST.  But for purposes of scoping the ground-based follow-up effort,
early elimination of giants is relatively unimportant.

\section{Masses of Super Earths}

Perhaps the most important promise of the TESS mission is that the
transiting planets it can find will be orbiting the nearest and
brightest stars of a given type. These will thus be the best targets
for follow-up observations, such as spectroscopy of planetary
atmospheres during transits and secondary eclipses with the James Webb
Space Telescope (JWST).  Furthermore, TESS will have the sensitivity
to reach down to Super Earths and even to a few Earth-sized planets
orbiting small stars.  The critical step for confirming the smallest
planets will be Doppler spectroscopy at the highest possible precision
in order to determine planetary masses from the spectroscopic orbits
of the host stars.  If radii and masses can be determined with
accuracies at the level of 5 and 10\%, theoretical models suggest that
it will be possible to deduce the structure and composition of Super
Earths, namely planets with a family resemblance to Earth, for example
to distinguish between desert rocky planets and water worlds.
Measuring masses for even the nearest and brightest examples of Super
Earths will push the limits of the present state-of-the art
spectrometers, such as HARPS on the 3.6-m telescope at the European
Southern Observatory and HIRES on Keck 1, which can now reach 1 \ms\ on
slowly-rotating and inactive solar-type stars.  However, many
observations are needed to average out astrophysical effects, and mass
determinations for transiting Super Earths will be very expensive in
telescope time.  For example, for an allocation of 75 nights on HARPS
and a strategy that focuses on deriving masses for the most
interesting small planets, we estimate a yield of perhaps two or three
dozen mass determinations.  The exciting prospect is that the best of
these planets may enable the iconic detection with JWST of
biologically interesting molecules in the atmosphere of a habitable
planet.

We are grateful to the Vulcan, TrES, and HAT extrasolar planet survey teams
for providing us with nearly 1000 candidate transiting planet systems.
Their work has led to the discovery of many transiting planets, has 
provided the training set for development of our follow-up procedures,
and is the basis for most of the numerical estimates in \S 5.
DWL thanks the Kepler Mission for partial support of this work through
NASA Cooperative Agreement NCC2-1390.

\clearpage

\end{document}